\documentclass[preprint,showpacs,preprintnumbers,pre,superscriptaddress]{revtex4-1}
\usepackage{graphicx}
\begin{document}

\title{Coefficient of performance at maximum $\chi$-criterion for Feynman ratchet as a refrigerator}
\author{Shiqi Sheng}
\affiliation{Department of Physics, Beijing Normal University, Beijing 100875, China}
\author{Pan Yang}
\affiliation{Department of Physics, Beijing Normal University, Beijing 100875, China}
\author{Z. C. Tu}\email[Corresponding author. Email: ]{tuzc@bnu.edu.cn}
\affiliation{Department of Physics, Beijing Normal University, Beijing 100875, China}
\affiliation{Kavli Institute for Theoretical Physics China, CAS, Beijing 100190, China}

%\date{\today}

\begin{abstract}The $\chi$-criterion is defined as the product of the energy conversion efficiency and the heat absorbed per unit time by the working substance [de Tom\'{a}s~\emph{et al.}, \textit{Phys. Rev. E}, \textbf{85} (2012) 010104(R)]. The $\chi$-criterion for Feynman ratchet as a refrigerator operating between two heat baths is optimized. Asymptotic solutions of the coefficient of performance at maximum $\chi$-criterion for Feynman ratchet are investigated at both large and small temperature difference. An interpolation formula, which fits the numerical solution very well, is proposed. Besides, the sufficient condition for the universality of  the coefficient of performance at maximum $\chi$ is investigated.
\end{abstract}
\maketitle

\section{Introduction}

One of the key topics in the finite-time thermodynamics is the efficiency at maximum power output (EMP) for heat engines. Since the pioneer work made by Curzon and Ahlborn~\cite{Curzon1975}, this topic has been fully investigated by many researchers~\cite{Andresen1977,Chen1989,ChenJC94,Bejan96,ChenL99,vdbrk2005,dcisbj2007,Schmiedl2008,Tu2008,Esposito2009,Apertet12,Esposito2010,Seifert12rev,GaveauPRL10,WangTu2011,wangtu2012,Izumida2012,WangHe,wangheinter,Guochen2013,wangtu13ctp,Rubin1979,Salamon1980,Tusheng13,Tusheng14,Hejizhou2013,Tu2012,ChengXT2013,Bao1999,Parrondo1996,ApertetPRE031161,ApertetPRE022137,ChenJCEPJB06,ChenJCMPLB10}. Recent researches are mainly focused on two issues: One is the universal EMP for tight-coupling heat engines operating between two heat baths at small temperature difference~\cite{vdbrk2005,dcisbj2007,Schmiedl2008,Tu2008,Esposito2009,Apertet12,Seifert12rev,ApertetPRE031161,ChenJCEPJB06,ChenJCMPLB10}; the other is the global bounds of EMP for heat engines operating between two heat baths~\cite{Esposito2010,GaveauPRL10,WangTu2011,wangtu2012,Izumida2012,WangHe,wangheinter,Guochen2013,wangtu13ctp,Tusheng13,Tusheng14}.

On the other hand, the optimal performance of refrigerators has also attracted much attention~\cite{Lingenchen1995,ChenDingsun11,Jincan1998,Jincanjpa09,Jizhouhe02,dcisbj2006,TRHWT2013,Mahler2010,Velasco1997,YanChen1990,RocoPRE12,WLTHRpre12,Izumida13,ApertetEPL40001,HeJZPRE13}. Recently, de Tom\'{a}s \emph{et al.} introduced a unified optimization $\chi$-criterion for heat devices including heat engines and refrigerators~\cite{RocoPRE12}. This $\chi$-criterion is defined as the product of the energy conversion efficiency and the heat absorbed by the working substance per unit time. The coefficient of performance (COP) at maximum $\chi$ was proved to be $\varepsilon_{CY}\equiv \sqrt{\varepsilon_{C}+1}-1$ for symmetric low-dissipation refrigerator, which is precisely the same with that derived by Yan and Chen~\cite{YanChen1990} for endoreversible refrigerator, where $\varepsilon_{C}$ is the Carnot efficiency for refrigerators.

Feynman introduced an imaginary ratchet device, which can work as heat engine as well as refrigerator, in his famous lectures~\cite{Feynmanbook}. Many discussions have been made on this device~\cite{Tu2008,ChenDingsun11,Jincanjpa09,ChenJCEPJB06,ChenJCMPLB10,HeJZPRE13}. The EMP for Feynman ratchet as a heat engine has been discussed in Ref.~\cite{Tu2008,ChenJCEPJB06,ChenJCMPLB10}. In this work, as a counter part, we investigate the COP at maximum $\chi$-criterion for Feynman ratchet as a refrigerator. Firstly, we introduce the Feynman ratchet as a typical tight-coupling refrigerator and optimize its $\chi$-criterion. We find its COP at maximum $\chi$ approaches to $\sqrt{\varepsilon_{C}}$ when the temperature difference between two heat baths is small. This asymptotic behavior is in consistent with the results derived by the present authors in Ref.~\cite{Tusheng13,Tusheng14}. Secondly, an interpolation formula, which fits the numerical solution well, is proposed. We also prove the relative error between the interpolation formula and the exact solution to be less than $0.8\%$. Finally, by constructing the mapping from Feynman ratchet as a tight-coupling refrigerator into the refined generic model proposed in~\cite{Tusheng14}, we investigate the sufficient conditions for the universality of $\varepsilon_{\ast}\rightarrow\sqrt{\varepsilon_{C}}$ at small temperature difference for refrigerators.

\section{Feynman ratchet as a refrigerator\label{sec-Feynmanref}}

\begin{figure}[htp!]\begin{center}
\includegraphics[width=8cm]{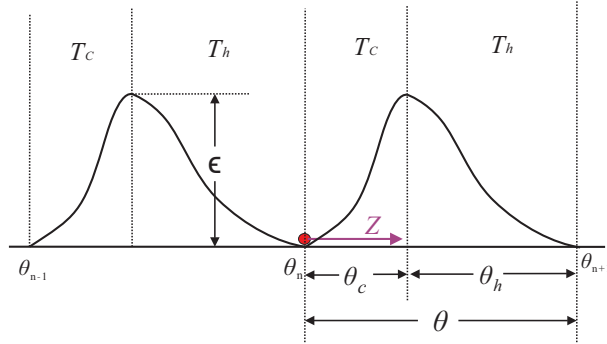} \caption{\label{fig1}Potential energy (the ordinate) as a function of the rotation angle (the abscissa) of the ratchet.}\end{center}
\end{figure}

The Feynman ratchet can be regarded as a B\"uttiker-Landauer model~\cite{Feynmanbook,Buttiker,Landauer}, i.e., a Brownian particle walking in a periodic lattice labeled by $\Theta_n,~(n=\cdots,-2,-1,0,1,2,\cdots)$ with a fixed step size $\theta$. The ratchet potential is schematically depicted in Fig.~\ref{fig1}, where the energy scale and the position of potential barrier are respectively denoted by $\epsilon$ and $\theta_c$ (or $\theta_{h}$). The Brownian particle is in contact with a cold bath at temperature $T_c$ in the left side of each potential barrier while it is in contact with a hot bath at temperature $T_h$ ($>T_c$) in the right side of each barrier. The particle is pulled from the left to the right side by a moment $z$ due to the external force. In steady state and overdamping condition, the forward and backward jumping rates can be respectively expressed as $\omega_{+}=k_{0}\mathrm{e}^{-(\epsilon-z\theta_c)/T_c}$ and $\omega_{-}=k_{0}\mathrm{e}^{-(\epsilon+z\theta_h)/T_h}$ according to the Arrhenius law \cite{Feynmanbook}. In the expressions of jumping rates, the Boltzmann factor is set to be $1$ while $k_{0}$ represents the bare rate constant with dimension of time$^{-1}$. For simplicity, we introduce two abbreviated notations $q\equiv \epsilon-z\theta_c$ and $w\equiv z\theta$.

For the relative large load $z$, the forward jumping rate can be larger than the backward one. In this case, the net current
\begin{equation}r\equiv\omega_{+}-\omega_{-}=k_{0}\left[\mathrm{e}^{-q/T_c}-\mathrm{e}^{-(q+w)/T_h}\right]\label{fluxfr}\end{equation}
is positive. In each forward step, the particle absorbs heat $q\equiv \epsilon-z\theta_c$ from the cold bath. Combining the input work $w\equiv z\theta$ done by the external load, the absorbed heat will be released into the hot bath when the particle jumps over the barrier. Thus the total heat $q+w$ will be released into the hot bath in each forward step. The energy conversion in each backward step is exactly opposite of that in forward step mentioned above. Thus the net power input can be expressed as $\dot{W}=wr$ while the heat absorbed from the cold bath or released into the hot bath per unit time can be expressed as $\dot{Q}_{c}=qr$ or $\dot{Q}_{h}=(q+w)r$, respectively. Obviously, when $r>0$ the heat flows from the cold bath to the hot one, and the power input (i.e., the mechanical flux) is proportional to the thermal fluxes ($\dot{Q}_{c}$ and $\dot{Q}_{h})$. It is in this sense that the Feynman ratchet is regarded as a tight-coupling refrigerator.

\section{COP at maximum $\chi$-criterion\label{sec-MaxFeynmanref}}
In this section, the $\chi$-criterion for Feynman ratchet as a refrigerator is optimized with respect to both internal and external parameters. The asymptotic solutions for COP at maximum $\chi$ are investigated and an interpolation formula is proposed.

\subsection{Maximization of the $\chi$-criterion\label{sec-MaxFeynmanref-A}}
The COP of Feynman ratchet can be expressed as
\begin{equation}\varepsilon\equiv \dot{Q}_{c}/\dot{W} =q/w.\label{eq-effi}\end{equation}
Simultaneously, the $\chi$-criterion can be expressed as
\begin{equation}\chi\equiv\varepsilon\dot{Q}_{c}=\frac{k_{0}q^{2}}{w}\left[\mathrm{e}^{-q/T_c}-\mathrm{e}^{-(q+w)/T_h}\right].\label{eq-chi}\end{equation}

Maximizing $\chi$ with respect to the internal barrier hight $\epsilon$ and the external load $z$, we can obtain
\begin{eqnarray}
       (2-q/T_c) \mathrm{e}^{-q/T_c}&=&(2-q/T_h)\mathrm{e}^{-(q+w)/T_h}, \label{MAXIMUM1}\\
        \mathrm{e}^{-q/T_c}&=& (1+w/T_h)\mathrm{e}^{-(q+w)/T_h}.\label{MAXIMUM2}
\end{eqnarray}
Combining Eq.~(\ref{eq-effi}) and the above two equations, we derive that the COP at maximum $\chi$-criterion ($\varepsilon_{\ast}$) satisfies the following transcendental equation:
\begin{equation}\label{eq-copmxchi}
\frac{\varepsilon_{C}-\varepsilon_{\ast}}{\varepsilon_{C}+1}\left(\frac{2}{\varepsilon_\ast}-\frac{1}{\varepsilon_{C}}\right)=\ln\frac{(2+\varepsilon_\ast)\varepsilon_{C}}{\varepsilon_\ast(\varepsilon_{C}+1)},
\end{equation}
where $\varepsilon_{C}=T_{c}/(T_{h}-T_{c})$ is the COP of Carnot refrigerators. Considering we do not optimize $\chi$ with respect to $\theta_{c}$ and $\theta_{h}$, it is interesting and surprising that $\varepsilon_\ast$ depends merely on $\varepsilon_{C}$ (or equivalently speaking, the relative temperature difference between two baths) rather than the position of the potential barrier $\theta_c$ even though the expressions of jumping rates contain $\theta_c$.

\subsection{Asymptotic solutions}
Since the analytic solution to Eq.~(\ref{eq-copmxchi}) cannot be achieved, we will investigate the asymptotic behaviors at large temperature difference ($\varepsilon_{C}\rightarrow 0$) and small temperature difference ($\varepsilon_{C}\rightarrow \infty$), respectively. In the former case, considering $0\le\varepsilon_\ast\le\varepsilon_{C}$, we transform Eq.~(\ref{eq-copmxchi}) into $2\varepsilon_{C}/\varepsilon_\ast+\varepsilon_\ast/\varepsilon_{C}-3=\ln(2\varepsilon_{C}/\varepsilon_\ast)$,
from which we obtain
\begin{equation}\varepsilon_\ast\simeq 0.524 \varepsilon_{C},\label{eq-coplim0}\end{equation}
when $\varepsilon_{C}\rightarrow 0$.
This behavior is different from that of $\varepsilon_{CY}\equiv\sqrt{\varepsilon_C+1}-1$ which gives $\varepsilon_{CA}=0.5\varepsilon_{C}$ when $\varepsilon_{C}\rightarrow 0$. Thus, except for $\varepsilon_\ast\rightarrow 0$, there is no universal behavior of COP at maximum $\chi$-criterion for refrigerators at large temperature difference.

On the other hand, it is not hard to prove $\varepsilon_\ast\rightarrow \infty$ and $\varepsilon_\ast/\varepsilon_{C} \rightarrow 0$ when the temperature difference between two baths is very small (i.e., $\varepsilon_{C}\rightarrow \infty$) by using the reduction to absurdity.
Since $0<\varepsilon_{\ast}<\varepsilon_{C}$, we assume that $\varepsilon_\ast$
approaches to a finite value $p$ between 0 and $\varepsilon_{C}$ when $\varepsilon_{C}\rightarrow\infty$. In this case, $1/\varepsilon_{C}\rightarrow 0$, $\varepsilon_\ast/\varepsilon_{C}=p/\varepsilon_{C}\rightarrow 0$. Then Eq.~(\ref{eq-copmxchi}) is transformed into $2/p=\ln(1+2/p)$ which cannot hold because $\ln(1+2/p)$ is always smaller than $2/p$ for any finite $p>0$. This contradiction implies that $\varepsilon_\ast\rightarrow \infty$ when $\varepsilon_{C}\rightarrow\infty$.
Similarly, since $0<\varepsilon_{\ast}<\varepsilon_{C}$, we assume that $\varepsilon_\ast/\varepsilon_{C}$ takes a finite value $p^\prime$ between 0 and 1 when $\varepsilon_{C}\rightarrow\infty$.
Then Eq.~(\ref{eq-copmxchi}) is transformed into
${2}/{p^\prime}-3+p^\prime=(1+\varepsilon_{C})\ln[(1+2/\varepsilon_{C}p^\prime)/(1+1/\varepsilon_{C})]\simeq {2}/{p^\prime} -1$.
From this equation we obtain $p^\prime =2$ which contradicts with the assumption $0<p^\prime<1$. This contradiction implies that $\varepsilon_\ast/\varepsilon_{C}\rightarrow 0$ when $\varepsilon_{C}\rightarrow\infty$.

If multiplying $1+\varepsilon_{C}^{-1}$ on both sides of Eq.~(\ref{eq-copmxchi}) and then
expanding it into a series of $\varepsilon_{\ast}^{-1}$ and $\varepsilon_{C}^{-1}$, we can derive
\begin{equation}{2}/{\varepsilon_{C}}-{\varepsilon_{\ast}}/{\varepsilon_{C}^2}+{2}/{\varepsilon_{\ast}\varepsilon_{C}}-{2}/{\varepsilon_{\ast}^2}=0,\label{eq-copmxchi2}\end{equation}
when we neglect the contribution of higher order terms. The terms
$\varepsilon_\ast/\varepsilon_{C}^2$ and $2/\varepsilon_\ast\varepsilon_{C}$ are of higher order relative to $1/\varepsilon_{C}$ since $\varepsilon_\ast\rightarrow \infty$ and $\varepsilon_\ast/\varepsilon_{C} \rightarrow 0$ when $\varepsilon_{C}\rightarrow \infty$, thus Eq.~(\ref{eq-copmxchi2}) is further simplified into $2/\varepsilon_{C}-2/\varepsilon_{\ast}^2\simeq 0$ when $\varepsilon_{C}\rightarrow \infty$. Its solution is
\begin{equation}\varepsilon_{\ast}\simeq \sqrt{\varepsilon_{C}},\label{eq-copsmalt}\end{equation}
which gives the asymptotic behavior of COP at maximum $\chi$-criterion for the Feynman ratchet as a refrigerator operating between two heat baths at small temperature difference.

\subsection{Interpolation formula}
\begin{figure}[htp!]\begin{center}
\includegraphics[width=8cm]{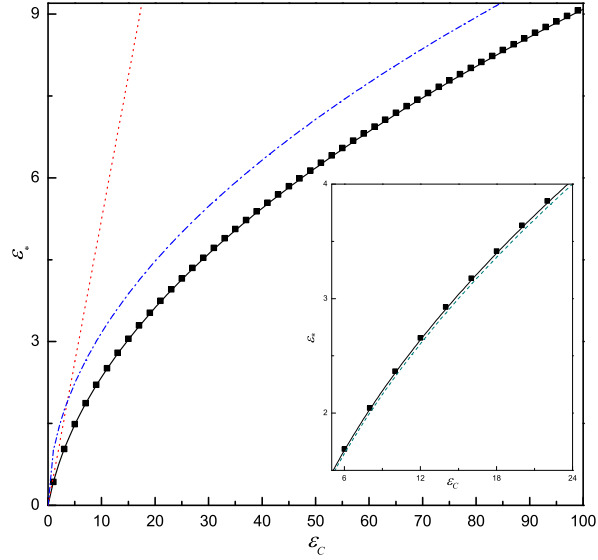}\caption{(Color online) COP at maximum $\chi$-criterion for the Feynman ratchet as a refrigerator. The numerical solutions to Eq.~(\ref{eq-copmxchi}) and interpolation formula (\ref{eq-intloat}) are depicted as the squares and solid line, respectively. The diagrams of functions (\ref{eq-coplim0}) and (\ref{eq-copsmalt}) are depicted as the dot line and dash dot line, respectively. Inset graph shows the diagram of interpolation formula (solid line) and that of $\varepsilon_{CY}\equiv\sqrt{\varepsilon_{C}+1}-1$ (dash line) as well as the numerical solutions (squares) in the range of $5<\varepsilon_{C}<24$.\label{fig2}}\end{center}
\end{figure}

We suggest using an interpolation formula
\begin{equation}\varepsilon_\ast=\sqrt{\varepsilon_{C}+\alpha^2}-\alpha,\label{eq-intloat}\end{equation}
with $\alpha=1/(2\times 0.524)=0.954$ as the approximate solution to Eq.~(\ref{eq-copmxchi}). It is easy to see that this formula degenerates into Eq.~(\ref{eq-coplim0}) (or Eq.~(\ref{eq-copsmalt})) when
$\varepsilon_{C}\rightarrow 0$ (or $\varepsilon_{C}\rightarrow \infty$), respectively.
We compare this interpolation formula with the numerical solution to Eq.~(\ref{eq-copmxchi}) shown in Fig.~\ref{fig2}.
Surprisingly, this interpolation formula (solid line) does extremely approach to the numerical solutions (squares) to Eq.~(\ref{eq-copmxchi}) obtained from the high precision computation when $\varepsilon_{C}$ takes values in a relatively large range. This formula fits the numerical data better than formulas (\ref{eq-coplim0}) and (\ref{eq-copsmalt})
depicted respectively as dot line and dash dot line in Fig.~\ref{fig2}. There exists a little difference between this interpolation formula and $\varepsilon_{CY}\equiv\sqrt{\varepsilon_{C}+1}-1$. Both of them are respectively depicted as the solid line and dash line in the inset of Fig.~\ref{fig2}, which reveals that the interpolation formula is much closer to the numerical solutions than $\varepsilon_{CY}\equiv\sqrt{\varepsilon_{C}+1}-1$.

Now we will investigate the relative error between this interpolation formula and the exact solution to Eq.~(\ref{eq-copmxchi}) although we cannot analytically achieve the exact solution in practice.

Introducing a variable $y\equiv \varepsilon_\ast/\varepsilon_{C}$, we transform Eq.~(\ref{eq-copmxchi})
into
\begin{equation}(1+\varepsilon_{C})[\ln(2+y\varepsilon_{C})-\ln y]+3-2/y-y =(1+\varepsilon_{C})\ln(1+\varepsilon_{C}).\end{equation}
Define two functions:
\begin{eqnarray}f(y,\varepsilon_{C})&\equiv&(1+\varepsilon_{C})[\ln(2+y\varepsilon_{C})-\ln y]+3-2/y-y,\nonumber\\
g(\varepsilon_{C})&\equiv&(1+\varepsilon_{C})\ln(1+\varepsilon_{C}).
\label{Feyref-12}\end{eqnarray}

Assume that $\bar{y}$ and $\tilde{y}$ represent the approximate solution and exact solution to equation $f(y,\varepsilon_{C})=g(\varepsilon_{C})$, respectively. Now we will calculate $f(\bar{y},\varepsilon_{C})-f(\tilde{y},\varepsilon_{C})$. On the one hand
\begin{equation}f(\bar{y},\varepsilon_{C})-f(\tilde{y},\varepsilon_{C})=f(\bar{y},\varepsilon_{C})-g(\varepsilon_{C})\label{eq-erro1}\end{equation}
since the exact solution $\tilde{y}$ satisfies $f(\tilde{y},\varepsilon_{C})=g(\varepsilon_{C})$. On the other hand,
\begin{equation}f(\bar{y},\varepsilon_{C})-f(\tilde{y},\varepsilon_{C})=\left(\frac{\partial f}{\partial y}\right)_{y=\bar{y}}(\bar{y}-\tilde{y}).\label{eq-erro2}\end{equation}
Combining Eqs.~(\ref{eq-erro1}) and (\ref{eq-erro2}), we derive the relative error
\begin{eqnarray}&&\left|\frac{\bar{y}-\tilde{y}}{\bar{y}}\right|=\frac{|f(\bar{y},\varepsilon_{C})-g(\varepsilon_{C})|}{\left|\bar{y}(\partial f/\partial y)_{y=\bar{y}}\right|}\label{Feyref-15}\\
&&=\frac{\left|(1+\varepsilon_C)\ln[(2+\bar{y}\varepsilon_C)/\bar{y}(1+\varepsilon_C)]+3-2/\bar{y}-\bar{y}\right|}{\left|2/\bar{y}-\bar{y}-2(1+\varepsilon_C)/(2+\bar{y}\varepsilon_C)\right|}\nonumber.
\end{eqnarray}

\begin{figure}[htp!]\begin{center}
\includegraphics[width=8cm]{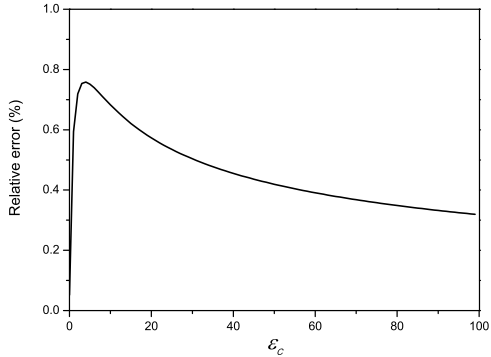}\caption{Relative error between the interpolation formula $\varepsilon_\ast=\sqrt{\varepsilon_{C}+\alpha^2}-\alpha$ and the exact solution to  Eq.~(\ref{eq-copmxchi}).}\label{figs-error}\end{center}
\end{figure}

With the consideration of interpolation formula $\varepsilon_\ast=\sqrt{\varepsilon_{C}+\alpha^2}-\alpha$ where $\alpha=0.954$, we have $\bar{y}\equiv\varepsilon_\ast/\varepsilon_{C}=\sqrt{1/\varepsilon_C+\alpha^2/\varepsilon_C^2}-\alpha/\varepsilon_C$.
Substituting it into Eq.~(\ref{Feyref-15}), we can estimate the relative error between interpolation formula $\varepsilon_\ast=\sqrt{\varepsilon_{C}+\alpha^2}-\alpha$ and the exact solution to Eq.~(\ref{eq-copmxchi}). The detailed results are shown in Fig.~\ref{figs-error} which reveals that the relative error is smaller than 0.8\%.

\section{Sufficient conditions of the Universality\label{sec-Weighted}}
The asymptotic behavior $\varepsilon_{\ast}\rightarrow \sqrt{\varepsilon_{C}}$ at small temperature difference for Feynman ratchet as a refrigerator is also shared by $\varepsilon_{CY}\equiv \sqrt{\varepsilon_{C}+1}-1$ and is in consistent with the results in Ref.~\cite{Tusheng13,Tusheng14} obtained from a refined generic model of refrigerator. This consistency prompt us to map Feynman ratchet as a refrigerator into the refined generic model of refrigerator and search for the sufficient conditions of the universality of $\sqrt{\varepsilon_{C}}$. Before this, we have to make a review of the refined generic model of refrigerator in detail, for the depiction in our previous work is oversimplified. We have to emphasize that although our discussions mainly focus on autonomous refrigerators since the Feynman ratchet is an autonomous engine, the refined generic model can also be applied to cyclic refrigerators by a different adoption of mechanical flux and force as discussed in Ref.~\cite{Tusheng14}.

\begin{figure}[htp!]\begin{center}
\includegraphics[width=8cm]{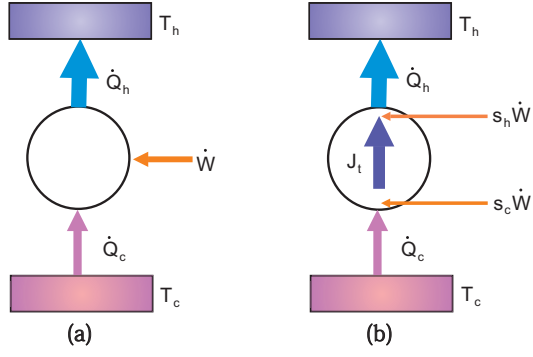}\caption{(Color online) Generic model of a refrigerator: (a) the convential version and (b) the refined version.}\label{fig4}\end{center}
\end{figure}

\subsection{Refined generic model of a tight-coupling refrigerator}
Fig.~\ref{fig4}(a) shows a conventional generic setup of Carnot-like refrigerator. In unit time, the working substance absorbs heat $\dot{Q}_{c}$ from the cold bath at temperature $T_{c}$ and releases heat $\dot{Q}_{h}$ into the hot bath at temperature $T_{h}$, while the amount of work $\dot{W}$ is done to the working substance by the environment. The dot denotes derivative with respect to time.

For an autonomous refrigerator in the steady state, the entropy production rate $\sigma$ can be expressed as
\begin{equation}\sigma =\beta _{h}\dot{Q}_{h}- \beta _{c}\dot{Q}_{c},\label{weighted-1}\end{equation}
where $\beta_{h}$ and $\beta_{c}$ denote the reciprocal of $T_{h}$ and $T_{c}$, respectively.

Following the procedure we have done to heat engines in Ref.~\cite{Tusheng14}, Eq.~(\ref{weighted-1}) can be transformed into
\begin{equation}\sigma =\beta _{c}\dot{W}+\dot{Q}_{h}\left( \beta _{h}-\beta _{c}\right),\label{weighted-2}\end{equation} or
\begin{equation}\sigma =\beta _{h}\dot{W}+\dot{Q}_{c}\left( \beta _{h}-\beta _{c}\right),\label{weighted-3}\end{equation}
with consideration of $\dot{Q}_{h}=\dot{Q}_{c}+\dot{W}$. Introduce two nonnegative weighted parameters $s_{c}$ and $s_{h}$ with $s_{c}+s_{h}=1$. Multiply Eqs.~(\ref{weighted-2})~and~(\ref{weighted-3}) by $s_{c}$ and $s_{h}$, respectively. Then the sum of both products lead to
\begin{equation}\sigma =(s_{c}\beta _{c}+s_{h}\beta_{h}) \dot{W}+(s_{c}\dot{Q}_{h}+ s_{h}\dot{Q}_{c})(\beta _{h}-\beta_{c}). \label{weighted-4}\end{equation}
This expression enlighten us to take the thermal flux $J_{t}$ as
\begin{equation} J_{t}\equiv s_{c}\dot{Q}_{h}+s_{h}\dot{Q}_{c}, \label{weighted-5}\end{equation}
with the conjugating thermal force
\begin{equation} X_{t}\equiv\beta_{h}-\beta_{c}.\label{weighted-6}\end{equation}
For the autonomous engine operating in the steady state, the power input $\dot{W}$ can be expressed as $\dot{W}=rW$, where $r$ is the net rate and $W$ is the elementary work in each mechanical step. Then the mechanical flux and force, which are related to the mechanical process, can be expressed as
\begin{equation}J_m\equiv r,\mathrm{~and~} X_m\equiv \beta W,\label{weighted-9}\end{equation}
respectively. $\beta$ denotes the weighted reciprocal of temperature
\begin{equation}\beta=s_{c}\beta _{c}+s_{h}\beta_{h}.\label{weighted-8}\end{equation}
With consideration of above definitions~(\ref{weighted-5})--(\ref{weighted-8}), we can exactly transform Eq.~(\ref{weighted-4}), the entropy production rate, into a canonical form
\begin{equation} \sigma=J_{m}X_{m}+J_{t}X_{t}. \label{weighted-sigma}\end{equation}
Besides, we can derive the power input
\begin{equation} \dot{W}=\beta^{-1}J_{m}X_{m}. \label{weighted-10}\end{equation}
From Eqs.~(\ref{weighted-5}),~(\ref{weighted-10}) and $\dot{Q}_{h}=\dot{Q}_{c}+\dot{W}$, $\dot{Q}_{c}$ and $\dot{Q}_{h}$ can be further expressed as
\begin{equation}\dot{Q}_{h}=J_{t}+s_{h}\dot{W},~\dot{Q}_{c}=J_{t}-s_{c}\dot{W}, \label{weighted-11}\end{equation}
by which we can revise the conventional generic model shown in Fig.~\ref{fig4}(a) into a refined version shown in Fig~\ref{fig4}(b). From Eq.~(\ref{weighted-10}) it is easy to see that the leading term of $\dot{W}$ is a quadratic order for small relative forces. As explained in Ref.~\cite{Tusheng14} for heat engines, $J_{t}$ may be regarded as the common leading term shared by $\dot{Q}_{c}$ and $\dot{Q}_{h}$. We can further assume that $s_{c}$(or $s_{h}$) represent the degree of coupling between the refrigerator and the cold (or hot) bath.

According to the work by Van den Broeck in Ref.~\cite{vdbrk2005}, this generic model of refrigerator can be described by the linear constitutive relation between thermodynamic fluxes and forces
\begin{equation}J_{m}=L_{mm}X_{m}+L_{mt}X_{t},~  J_{t}=L_{tm}X_{m}+L_{tt}X_{t},\label{VDB-3}\end{equation}
where the Onsager coefficients satisfy  $L_{mm}\geq 0$, $L_{tt}\geq 0$, $L_{mm}L_{tt}-L_{mt}L_{tm}\geq 0$ and $L_{mt}=L_{tm}$.
Considering the tight-coupling condition $L_{mt}^{2}=L_{tm}^{2}=L_{mm}L_{tt}$, the thermal flux is proportional to the mechanical flux
\begin{equation}J_{t}/{J_{m}}=L_{mt}/{L_{mm}}=\xi.\label{VDB-4}\end{equation}

From Eq.~(\ref{weighted-11}) and the tight-coupling condition Eq.~(\ref{VDB-4}), the heat absorbed from the cold bath can be further expressed as
\begin{equation}\dot{Q}_{c}=(\xi -s_{c}\beta^{-1} X_{m})J_{m}. \label{MAX-1}\end{equation}
Substituting $\dot{Q}_{c}$ into Eq.~(\ref{eq-effi}), we have the expression of COP as
\begin{equation}\varepsilon \equiv \dot{Q}_{c}/\dot{W}= \xi\beta/X_{m} -s_{c}.\label{MAX-2}\end{equation}
Thus the target function $\chi\equiv\varepsilon\dot{Q}_{c}$ can be derived as following
\begin{equation}\chi=(\xi-s_{c}X_{m} /\beta)^2\beta J_{m} /X_{m}.\label{MAX-3}\end{equation}
Maximizing $\chi$ with respect to $X_{m}$ for given $T_{c}$ and $T_{h}$, we obtain
\begin{equation}\left(s_c+\frac{\beta \xi}{X_m}\right) \frac{J_m}{X_m} +\left(s_c -\frac{\beta \xi}{X_m}\right)\frac{\partial J_{m}}{\partial X_{m}}=0.\label{MAX-4}\end{equation}
With consideration of the linear constitutive relation Eq.~(\ref{VDB-3}) and tight-coupling condition Eq.~(\ref{VDB-4}), we can derive from Eq.(\ref{MAX-4})
\begin{equation}{\beta \xi}/{ X_{m}} = \sqrt{{s_{c}^2}/{4}-{2s_{c}\beta}/{X_{t}}}-{s_{c}}/{2}.\label{MAX-5}\end{equation}
Substituting Eq.~(\ref{MAX-5}) into Eq.~(\ref{MAX-2}), we obtain the COP at maximum $\chi$ to be
\begin{equation}\varepsilon _{\ast} =\sqrt{2s_{c}\varepsilon _{C}+9s_{c}^{2}/4}-3s_{c}/2, \label{MAX-6}\end{equation}
with consideration of Eq.~(\ref{weighted-6})~,~(\ref{weighted-8}) and $\varepsilon_{C}=T_{c}/(T_{h}-T_{c})$.

In particular, in the case of symmetric coupling, $s_{c}=s_{h}={1}/{2}$, Eq.~(\ref{MAX-6}) can be simplified into
\begin{equation}\varepsilon _{\ast} =\sqrt{\varepsilon _{C}+9/16}-3/4. \label{MAX-7}\end{equation}
It is easy to see from Eq.~(\ref{MAX-7}), if the symmetric coupling condition is satisfied, the COP at maximum $\chi$ for refrigerator approaches to $\sqrt{\varepsilon_{C}}$ when temperature difference is small ($\varepsilon_{C}\rightarrow\infty$).

\subsection{Mapping Feynman ratchet as a refrigerator into the refined generic model}
Considering the Feynman ratchet as a refrigerator shown in Fig.~\ref{fig1}, the heat absorbed from the cold bath and released into the hot bath per unit time can be expressed as
\begin{equation} \dot{Q}_{c}=\epsilon r-\frac{\theta_{c}}{\theta}z\theta r,~\dot{Q}_{h}=\epsilon r+\frac{\theta_{h}}{\theta}z\theta r, \label{Map3-1}\end{equation}
respectively. $r$ is the net current in Eq.~(\ref{fluxfr}). The power input can be expressed as $\dot{W}=z\theta r$. Comparing Eq.~(\ref{Map3-1})~with~(\ref{weighted-11}), we can straightforwardly derive the weighted thermal flux and the weighted parameters as
\begin{equation} J_{t}\equiv \epsilon r,\label{Map3-2}\end{equation}
and
\begin{equation} s_{c}=\frac{\theta _{c}}{\theta},~s_{h}=\frac{\theta _{h}}{\theta}, \label{Map3-3}\end{equation}
respectively. Obviously, the weighted parameters indeed reflect the degree of coupling strength between the particle and the cold bath or hot bath. From Eq.~(\ref{weighted-9}), we can get the mechanical flux and force
\begin{equation} J_{m}=r,~X_{m}=\beta z\theta,\label{Map3-4}\end{equation}
where $\beta$ satisfying Eq.~(\ref{weighted-8}).

It is easy to see that the weighted thermal flux $J_{t}$ is indeed tightly coupled with the mechanical flux $J_{m}$ with the scale factor $\xi=J_{t}/J_{m}=\epsilon$ from Eqs.~(\ref{Map3-2})~and~(\ref{Map3-4}). It should be emphasized that this relation may not be hold when we take the energy transaction due to kinetic energy into consideration. The specific discussions on this point are beyond the scope of our present work.

Substituting Eqs.~(\ref{Map3-2})--(\ref{Map3-4}) into Eq.~(\ref{fluxfr}), we have
\begin{eqnarray} J_{m}=k_{0}e^{-\beta \epsilon}[&&e^{s_{h}\epsilon X_{t}+s_{c}X_{m}-s_{c}s_{h}X_{m}X_{t}/{\beta}}\nonumber \\
&&-e^{-s_{c}\epsilon X_{t}-s_{h}X_{m}-s_{c}s_{h}X_{m}X_{t}/{\beta}}].\label{Map3-5}\end{eqnarray}
Under the linear approximation of Taylor expansion, $J_{m}$ can be further expressed as $J_{m}=k_{0}e^{-\beta \epsilon}\left(X_{m}+\epsilon X_{t}\right)$. Then the Onsager coefficients can be derived
\begin{eqnarray}
&&L_{mm}=k_{0}e^{-\beta \epsilon},\nonumber\\
&&L_{mt}=L_{tm}=k_{0}\epsilon e^{-\beta \epsilon},\label{Map3-6-onsager}\\
&&L_{tt}=k_{0}\epsilon^{2}e^{-\beta \epsilon}.\nonumber
\end{eqnarray}

Therefore, the Feynman ratchet as a refrigerator is mapped into the refined generic model approximately. The weighted parameters and Onsager coefficients satisfy Eqs.~(\ref{Map3-3})~and~(\ref{Map3-6-onsager}), respectively. Thus, symmetric coupling condition ($s_{c}=s_{h}=1/2$) is a sufficient condition of the COP at maximum $\chi$ approaches to $\sqrt{\varepsilon_{C}}$ for Feynman ratchet when $\varepsilon_{C}\rightarrow\infty$.

\subsection{Sufficient conditions}
If we take the second order term of $X_{m}$ and $X_{t}$ in Eq.~(\ref{Map3-5}) into consideration, $J_{m}$ can be expressed as $J_{m}=k_{0}e^{-\beta \epsilon}\left[\left(X_{m}+\epsilon X_{t}\right)+\frac{s_{c}-s_{h}}{2}\left(X_{m}^{2}-\epsilon ^{2}X_{t}^{2}\right)\right]$. Substituting $J_{m}$ into Eq.~(\ref{MAX-4}), we can derive the relationship between $X_{m}$ and $X_{t}$ at maximum $\chi$ as
\begin{eqnarray} &&\frac{X_{t}}{X_{m}}\left[1-\frac{1}{2}\left(s_{c}-s_{h}\right)\epsilon X_{t}\right]\left[s_{c}\epsilon+\frac{\beta \epsilon ^{2}}{X_{m}}\right]\nonumber\\
&&+\frac{3}{2}s_{c}(s_{c}-s_{h})X_{m}+2s_{c}-\frac{1}{2}\beta \epsilon (s_{c}-s_{h})=0.\label{Map3-7}\end{eqnarray}
Eq.~(\ref{MAX-5}) enlighten us assuming $\frac{1}{X_{m}}=\frac{A}{\sqrt{-X_{t}}}+B$ as the approximate solution of Eq.~({\ref{Map3-7}) when the temperature difference of the two heat baths is small ($X_{t}\rightarrow 0$). $A$ and $B$ are parameters independent of $X_{m}$ and $X_{t}$.
Substituting this trial solution into Eq.~(\ref{Map3-7}), we can derive
\begin{equation} -\beta \epsilon ^{2}A^{2}+2s_{c}-\frac{1}{2}(s_{c}-s_{h})\beta \epsilon =0,\label{Map3-8}\end{equation}
with consideration of the small relative temperature difference condition ($X_{t}\rightarrow 0$). From Eq.~(\ref{Map3-8}) we can derive the asymptotic expression of $A$ to be
\begin{equation} A=\sqrt{\frac{2s_{c}}{\beta \epsilon ^{2}}-\frac{s_{c}-s_{h}}{2\epsilon}},\label{Map3-9}\end{equation}
when $X_{t}\rightarrow 0$.

Considering Eqs.~(\ref{MAX-2}),~(\ref{Map3-9}) and $\varepsilon_{C}=T_{c}/(T_{h}-T_{c})$, we can derive the asymptotic COP at maximum $\chi$ for Feynman ratchet as a refrigerator to be
\begin{equation} \varepsilon_{\ast}=\sqrt{\left[2s_{c}-\frac{1}{2}(s_{c}-s_{h})\beta \epsilon\right] (\varepsilon_{C}+s_{c})}+\beta \epsilon B-s_{c}. \label{Map3-10}\end{equation}

From Eq.~(\ref{Map3-10}), we can derive that $\varepsilon_{\ast}\rightarrow \sqrt{\varepsilon_{C}}$ requires $s_{c}=s_{h}=1/2$ or $\beta\epsilon\rightarrow 2$. This alternative is in consistent with the statement that symmetric coupling condition is just the sufficient condition for $\varepsilon_{\ast}\rightarrow\sqrt{\varepsilon_{C}}$ when $\varepsilon_{C}\rightarrow\infty$ in Ref.~\cite{Tusheng14}.

Assume $\beta\epsilon =2+\varphi /\sqrt{\varepsilon_{C}}$ when $\varepsilon_{C}\rightarrow \infty $, where $\varphi$ is a finite constant. By considering $q=\epsilon -s_{c}w$ and $w$ to be the quadratic order term of $1/\varepsilon_{C}$, we have
\begin{equation} \beta q\rightarrow 2+\frac{\varphi}{\sqrt{\varepsilon_{c}}}+O(\frac{1}{\varepsilon_{C}}),\label{Map3-11}\end{equation}
where $O(\frac{1}{\varepsilon_{C}})$ denotes the higher order term of $1/\varepsilon_{C}$. Substituting (\ref{Map3-11}) into Eqs.~(\ref{MAXIMUM1})~and~(\ref{MAXIMUM2}), we can derive
\begin{eqnarray}&&-\frac{2}{s_{c}+\varepsilon_{C}}-\frac{\varphi}{\sqrt{\varepsilon_{C}}(s_{c}+\varepsilon_{C})}-\frac{2+\varphi/\sqrt{\varepsilon_{C}}}{\varphi\sqrt{\varepsilon_{C}} +2(1-s_{c})+\varphi/\sqrt{\varepsilon_{C}}}\nonumber \\
&&=\ln\left[1-\frac{2+\varphi/\sqrt{\varepsilon_{C}}}{\varphi\sqrt{\varepsilon_{C}} +2(1-s_{c})+\varphi/\sqrt{\varepsilon_{C}}}\right].\label{Map3-13}\end{eqnarray}
It is easy to see from Eq.~(\ref{Map3-13}) that this equality is true when $\varepsilon_{C}\rightarrow\infty$. In other words, when we adopt $\beta\epsilon =2+\varphi /\sqrt{\varepsilon_{C}}$, the COP at maximum $\chi$ for Feynman ratchet at small temperature difference indeed approaches to $\sqrt{\varepsilon_{C}}$ regardless of the value of $s_{c}$ and $s_{h}$. This result is in consistent with the discussion in Sec.~\ref{sec-MaxFeynmanref-A} that $\varepsilon_{\ast}$ is independent of $\theta_{c}$ and $\theta_{h}$.

\section{Conclusion and discussion\label{sec-discussion}}
In the above discussions, we investigate the COP at maximum $\chi$-criterion for the Feynman ratchet as a refrigerator and find that the corresponding COP can be approximately expressed as interpolation formula~(\ref{eq-intloat}). In the limit of small temperature difference between two heat baths, this formula and $\varepsilon_{CY}\equiv \sqrt{\varepsilon_{C}+1}-1$, which has been proved to be the COP at maximum $\chi$-criterion for endoreversible refrigerators~\cite{YanChen1990} and low-dissipation refrigerators~\cite{RocoPRE12}, share the same asymptotic behavior~(\ref{eq-copsmalt}). This universal asymptotic behavior, which is in consistent with the results obtained from a refined generic model of refrigerator proposed by the present authors~\cite{Tusheng14}, could be considered as the counterpart of universal EMP for tight-coupling heat engines in the presence of left-right symmetry~\cite{Esposito2009}.

Additionally, by mapping Feynman ratchet as a refrigerator into the refined generic model, we find that the asymptotic solution of the COP at maximum $\chi$ approaches to $\sqrt{\varepsilon_{C}}$ when $s_{c}=s_{h}=1/2$ or $\beta\epsilon =2+\varphi /\sqrt{\varepsilon_{C}}$ is satisfied. These two factors can be seen as the sufficient conditions of the universality of Eq.~(\ref{eq-copsmalt}) for Feynman ratchet as a refrigerator working at small temperature difference.

\section*{Acknowledgement}
The authors are grateful to the financial support from National Natural Science Foundation of China (Grant NO. 11322543).

\end{document}